\documentclass[twocolumn,floats,floatfix,superscriptaddress,aps,nofootinbib]{revtex4}
\usepackage{bm}
\usepackage{graphicx}
\usepackage{amsmath}
\usepackage{amsfonts}
\usepackage{mathrsfs}

\usepackage[utf8]{inputenc}
\usepackage{tabularx} 

\usepackage{xcolor}
\usepackage[normalem]{ulem}

\def\Lk{\hbox{\sl Lk}}
\def\Wr{\hbox{\sl Wr}}
\def\Tw{\hbox{\sl Tw}}
\begin{document}

\title{The Topology of Dislocations in Smectic Liquid Crystals}

\author{Randall D. Kamien}\email{kamien@upenn.edu}
\affiliation{Department of Physics and Astronomy, University of Pennsylvania, 209 South 33rd Street, Philadelphia, Pennsylvania 19104, USA}
\author{Ricardo A. Mosna}\email{mosna@ime.unicamp.br}
\affiliation{Departamento de Matem\'atica Aplicada, Universidade Estadual de Campinas, 13083-859, Campinas, SP, Brazil}
\affiliation{Department of Physics and Astronomy, University of Pennsylvania, 209 South 33rd Street, Philadelphia, Pennsylvania 19104, USA}

\begin{abstract}
The order parameter of the smectic liquid crystal phase is the same as that of a superfluid or superconductor, namely a complex scalar field.  We show that the essential difference in boundary conditions between these systems leads to a markedly different topological structure of the defects.  Screw and edge defects can be distinguished topologically.  This implies an invariant on an edge dislocation loop so that smectic defects can be topologically linked not unlike defects in ordered systems with non-Abelian fundamental groups.
\end{abstract}



\maketitle

\section{Introduction and Summary}

Classical dynamics is formulated through a collection of sometimes quite complex \cite{Einstein} and often tedious \cite{jackson} differential equations, a consequence of the implicit smooth structure of time evolution.  From this perspective, dynamics connects initial and final states through a homotopy in the configuration space of the system.  When that space has closed, non-contractible loops ({\sl i.e.}, an ``interesting'' topology), conservation laws ensue and non-trivial winding classes result in topological defects \cite{klemanmicheltoulouse,mermin79}.  In three dimensions, defect lines are characterized by the first homotopy group of the ground state manifold (GSM), $\pi_1(GSM)$.  In particular, when $\pi_1$ is non-Abelian, defect loops with homotopy classes $\alpha$ and $\beta$ are topologically linked, leaving behind a tether in class $\alpha\beta\alpha^{-1}\beta^{-1}$ whenever they are crossed \cite{poenaru77, toulouse,mineev,mermin79, michel80}.  It follows that the simple superfluid \cite{kosterlitz73} or its gauged cousin, the superconductor \cite{Abrikosov}, should not enjoy topologically tangled defects since their GSM is the circle ${\bf S}^1$ with the Abelian fundamental group $\mathbb{Z}$.  
\begin{figure}[b]
\centerline{\includegraphics[width=0.5\textwidth]{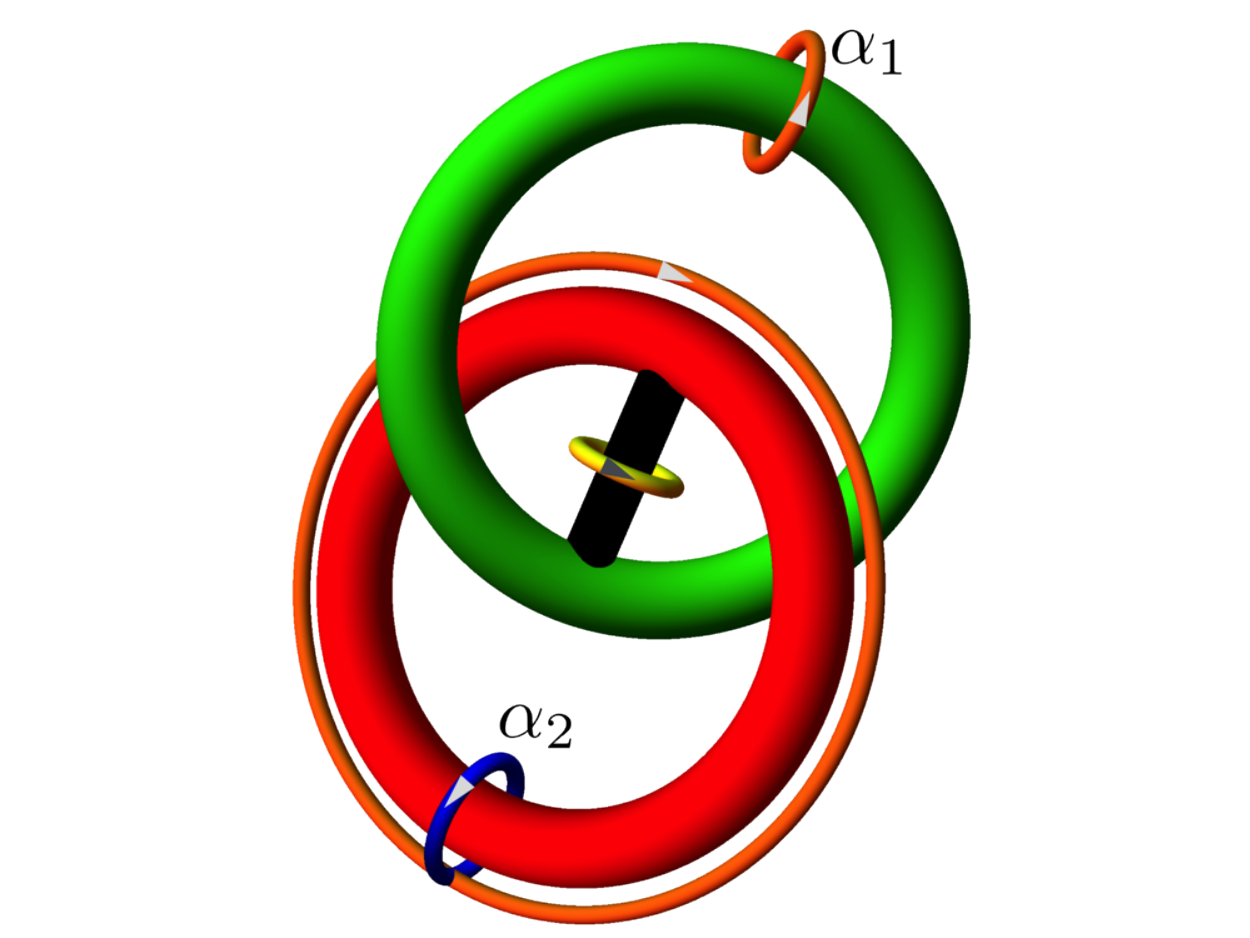}}
\caption{\label{fig:hopf} The red and green dislocation loops are linked and their equivalence classes $\alpha_1$ and $\alpha_2$ are measured by the orange and blue Burgers' circuits, respectively.  Note that because the red and green loops are linked, the orange loop around the ``equator'' of the red loop must measure the class from the green loop.  In Fig. 2 we unwrap the red torus to demonstrate that the measurement around the black tether (by the yellow circuit) is in the class $\alpha_1\alpha_2\alpha_1^{-1}\alpha_2^{-1}$.
}
\end{figure}

Liquid crystals provide a more complex GSM and, for instance, the biaxial nematic \cite{mermin79,poenaru77,cholsmec} is the ``poster child'' for non-Abelian defects.   In that system, topological defects are characterized by elements of a small but non-Abelian group leading to situations in which the commutator of elements is not the identity.  In comparison, the uniaxial nematic, upon which the smectic-A phase is based, is quite different.  Its line defects are characterized by $\mathbb{Z}_2$ and, as a result, all possible commutators yield the identity.
The topology of the smectic phase, however, is more subtle -- de Gennes drew a strong analogy between the smectic and the superconductor \cite{degennes}.  This powerful analogy led to a better understanding of the distinction between global and local symmetries \cite{Halsey}, the prediction of the analog of the Abrikosov phase \cite{Abrikosov} dubbed the TGB phase \cite{Renn}, and a dizzying number of different results on the critical behavior of the nematic to smectic-A transition \cite{TonerNelson, Toner, Lubensky, Patton}.  However, the smectic has both broken translation and rotational symmetry that spoils the standard homotopy description of defects which works perfectly for the superfluid \cite{mermin79,trebin82,klemanmichel,kurik88,chen09}.  Not unlike line and point  defects in nematics \cite{janich87,loops} the disclinations act upon the dislocations \cite{kurik88} creating ambiguities in free homotopy classes and, in the case of smectics, obstructions to generating the entire homotopy group \cite{poenaru,chen09}.   In the absence of disclinations, the smectic order parameter $\psi=\vert\psi\vert e^{i\phi}$ is precisely that of the superfluid or superconductor; we might expect that the topological classification of dislocations would be identical to that of vortices in superfluids.   In this paper, we will provide explicit constructions and general arguments to show that this is, in fact, not the case.  Dislocations in smectics can be topologically linked together despite the fact that the phase field $\phi\in{\bf S}^1$, suggesting a commuting set of topological defects.  Our analysis shows, moreover, that there is a discernible difference between edge and screw dislocations and suggests that the nature of the dislocation linking lies in the necessary decomposition of edge dislocations into disclination dipoles and their generalizations.  

In the next section we will provide a quick overview of the topological theory of defects with an emphasis on topological linking.  We will demonstrate why two linked topological defects sometimes require additional defects when the fundamental group, the group of closed loops around the ground state manifold (GSM), is non-Abelian.  We will contrast this with the geometric and energetic linking in the case of an Abelian fundamental group and set the appropriate stage to study topological linking.  In the third section we will define the smectic free energy and its required boundary conditions making constant contrast with the superfluid.    In particular, we will emphasize the difference in boundary conditions -- the superfluid has a phase field that goes to a constant on the boundary (or at infinity) while the smectic phase field must attain a constant value of $\nabla\phi$, representing the smectic ground state.  This difference is what changes the nature of linking, especially when we keep in mind that we must control the boundary conditions on the outer boundary of the sample in order for topological charge to be conserved -- we cannot let a defect escape through the surface and expect any of this to work!  Following that discussion we will show that these boundary conditions enable us to distinguish a screw dislocation from an edge dislocation, {\sl topologically}: the edge necessarily breaks up into a disclination pair while the screw does not.  This allows us to disentangle the geometric features of the defects from the topological ones.  Finally, using this insight in the fifth section, we will argue that dislocation loops in smectics can be topologically linked and provide an explicit linking invariant.   To complete the discussion and resolve the complexity of the overlapping textures of nearby defects, we need to establish a continuously parameterized set of related smectic states. We dub this new structure $q$-homotopy: $q$ is the wavenumber of the equilibrium layer spacing and we relate two different states with different values of $q$ but with the same set of defects with the same set of topological charges.   Before concluding, we outline another interpretation of the linking topology in terms of knots.  We end, as usual, with conclusions and future prospects.  {\sl Les jeux sont faits}.

\section{Overview of Topological Approach to Linking}
We start our discussion and set the stage by considering the state of the art.
Linking invariants for ${\bf S}^1$ ground state manifolds exist.  Recall  the helicity \cite{moffat}:
\begin{equation}
H = \int d^3\!x\,{\bf u}\cdot\left(\nabla\times{\bf u}\right)
\end{equation}
where ${\bf u}$ is the fluid velocity.  As Moffat noted, inviscid, incompressible Navier-Stokes flows characterized 
by $\partial_t {\bf u} + {\bf u}\cdot\nabla{\bf u} = -\nabla(p/\rho_0)$, conserve $H$.  In the case of a superfluid where ${\bf u}=\nabla\phi$, $H$ is the mutual linking number of all the defects when each defect has unit quantum of vorticity \cite{nelsonhw} -- in short, $\nabla\!\times\!{\bf u}$ points along the tangent to the vortices and is a delta function in the perpendicular plane.  This renders the volume integral a set of line integrals around the closed vortex loops and, upon integrating $\oint d\boldsymbol{\ell}\cdot {\bf u}$, one gets the vortex flux piercing the capping surface of each loop.  Since the flux is quantized, one might think that the linking number of vortex loops in a superfluid should be conserved for topological reasons. However, there is a self-interaction term that is highly sensitive to the shape of each loop and requires that each loop have a framing to track the helicity through inflections of the loops \cite{moffat2}.   Is their linking stable from a topological perspective? Not according to homotopy theory since $\pi_1({\bf S}^1)=\mathbb{Z}$ is Abelian \cite{poenaru77,mermin79}.  Recall, however, that even when the fundamental group of the GSM is non-Abelian we do not consider the crossing of defect lines and the unlinking of linked loops.  The difference lies in the tether that is left behind as we pull one loop past the other.  When $\pi_1(GSM)$ is Abelian the tether carries no net topological charge; when it is non-Abelian the tether has a charge and can be measured via a topological measurement around it.  The defects define the singular set of loops $\Sigma$ that is removed from the sample $M$ and we only consider maps from $M\setminus \Sigma$ to the GSM.  In both cases we do not  change the topology of the sample by crossing the singular sets of the defects and preserve the ambient volume around them.\footnote{The study of smooth maps between samples with {\sl different} topology is much more complex and should be the topic of an excellent review article.} So, tautologically, the linking number is conserved when we maintain $\Sigma$ up to ambient isotopy \cite{knots}.  In other words, even if we were to keep {\sl precisely} the same geometry of the dislocation loops, there would be a topological difference between the defects in a superfluid and a smectic.  The essential issue is that  boundary conditions force some defects to leave behind a tell-tale topological tether in the smectic.  
\begin{figure}
\centerline{\includegraphics[angle=270,origin=c,width=0.4\textwidth]{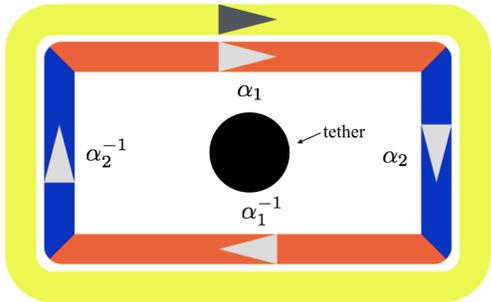}}
\caption{\label{fig:hopf2}  We can map the surface of the torus to the rectangle shown above.  If we measure the homotopy class of the boundary of the rectangle we get the commutator of $\alpha_1$ and $\alpha_2$.  If the commutator $\alpha_1\alpha_2\alpha_1^{-1}\alpha_2^{-1}\ne 1$  then there must be a defect somewhere on the surface -- the base of the black tether shown in Fig.~\ref{fig:hopf}.
}
\end{figure}

We start by reminding the reader about the nature of a non-Abelian $\pi_1(GSM)$ \cite{GPA}.  We closely follow the logic in Kl\'eman's early work \cite{klemanloops}.  Consider two dislocation loops in a Hopf link, as shown in Fig.~\ref{fig:hopf}.  We can measure the winding through the GSM for each loop by making a Burgers circuit around them.  The winding is characterized by an element of the equivalence class $\alpha\in\pi_1(GSM,x_0)$ of the fundamental group based at $x_0$. (In the Abelian case we may employ unbased homotopy). However, because these loops are embedded and linked, we know that there is another winding, around the ``equator'' of each loop as shown in Fig.~\ref{fig:hopf}.  If a loop is unlinked then we only get the trivial element around the equator.  

So what?  Consider a torus that surrounds each dislocation loop.  We use this for the measuring of these charges in $\pi_1(GSM)$, just as we use a circle in two dimensions or a sphere to measure $\pi_2(GSM)$ in three dimensions \cite{loops}.  Of course, we assume as usual that the map from the sample to the GSM is smooth on the torus, which is away from all defects.  However, we can unwrap the surface of the torus in the usual way where we use a rectangle with identified edges but now the edges carry the elements of the fundamental group.  Calculating the winding around the rectangle, as shown in Fig.~\ref{fig:hopf2}, we see that if the two charges are $\alpha_1$ and $\alpha_2$, the first homotopy class of the border, the yellow loop, must be $\alpha_1\alpha_2\alpha_1^{-1}\alpha_2^{-1}$, the commutator.  If the elements do not commute then there {\sl must be a defect somewhere on the surface} -- the tether is required by the topology and the embedding of the dislocation loops.    Were we to take two unlinked loops and try to link them we would not only encounter an energetic barrier associated with the defect cores, we would discover a topological obstruction as well -- the creation of a new topological defect, the tether.  

In superfluids and superconductors, there is an energetic barrier to crossing but no topological barrier.  In a biaxial nematic, on the other hand, the crossing is topologically forbidden.

\section{Smectics versus Superfluids}

The smectic is different because of its boundary conditions.  The free energy for a smectic has the following general form:
\begin{equation}\label{eq:fe}
F=\frac{1}{2}\int d^3\!x \left\{ B\left(\vert\nabla\phi\vert -\vert q\vert\right)^2 + K\left(\nabla^2\phi\right)^2\right\},
\end{equation}
where $2\pi/q$ is the ground state spacing, $B$ is the compression modulus, and $K$ is the bend modulus.  Any number of nonlinear embellishments \cite{geom} to (\ref{eq:fe}) are allowed as long as they all prefer the same ground state at infinity: equally spaced, flat layers.  Note that this ground state implies that on the boundary $\vert\nabla\phi\vert=q$ and $\phi$ is {\sl not constant} but is, rather, a linear function.   In the following we will, without loss of generality, take $\nabla\phi || \hat z$ as the boundary condition.  The level sets, $\phi\in 2\pi \mathbb{Z}$, locate the smectic layers -- it is equivalent and often useful to instead define the layers through the density variation  $\delta\rho\propto \cos(\phi)$ so that the ``peaks'' are defined by $\cos(\phi) =1$.  Though the material density variation  is single valued the phase field $\phi$ need not be: dislocations are characterized by their integer charge, $n\in\mathbb{Z}$, defined by $\oint_{\partial M} d{\boldsymbol{\ell}}\cdot\nabla\phi =2\pi n$.  Equivalently, the Burgers scalar $b=2\pi n/q$ is a multiple of the layer spacing and measures the net displacement around the Burgers circuit.  Note that our choice of boundary condition implies that there is no net dislocation charge in the sample.  This should be contrasted with the superfluid or superconductor.  When those systems have zero net defect charge, $\phi$ goes to a constant and the sample can be treated as ${\bf S}^2$ or ${\bf S}^3$ in two- or three-dimensions, respectively.  The smectic phase, on the other hand must have an accumulation of an infinite number of layers at infinity, leading to a defect in ${\bf S}^3$ that only vanishes when $q=0$.

\begin{figure}
\begin{center}
\includegraphics[width=0.5\textwidth]{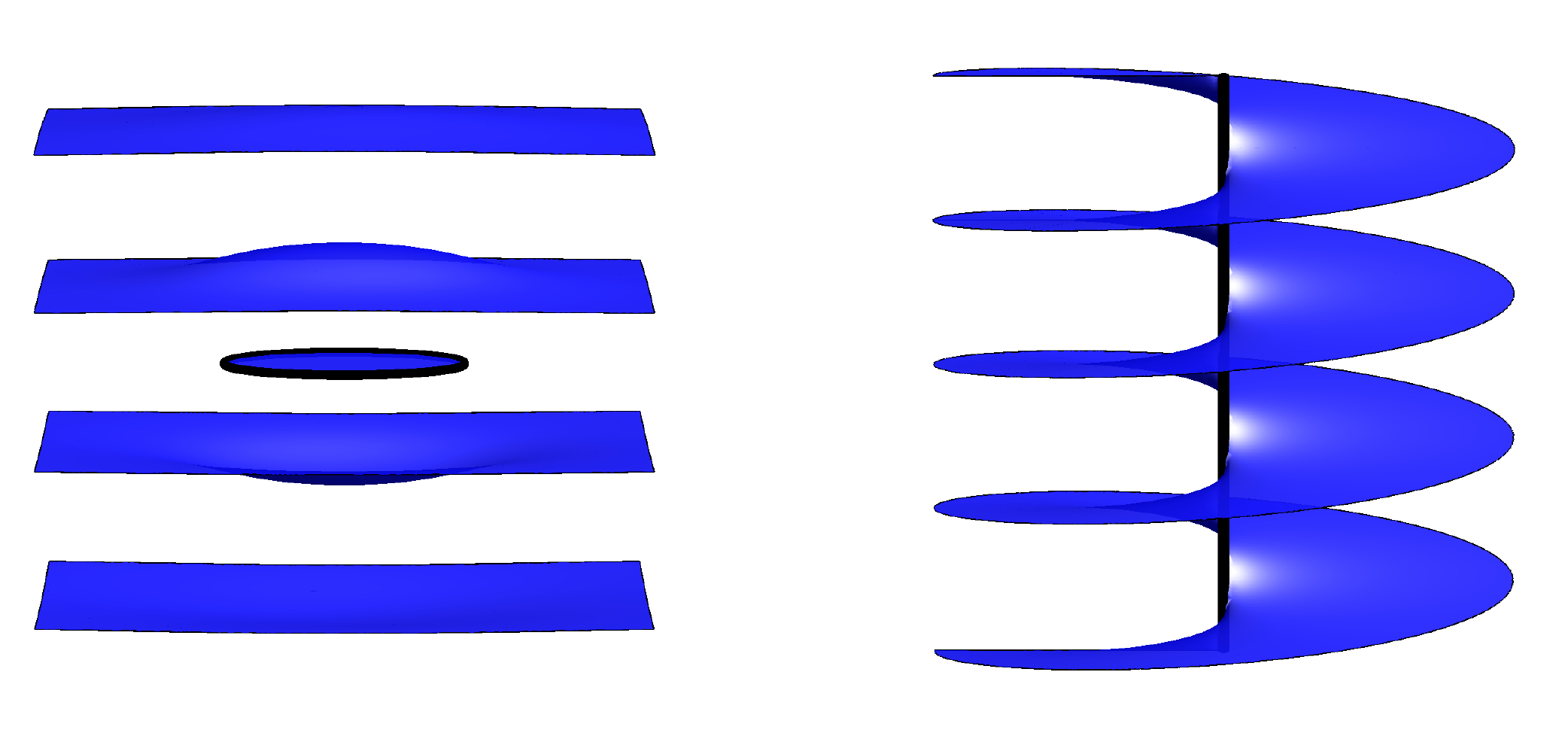}
\end{center}
\caption{\label{fig:pure_dislocations} The geometry of a pure edge dislocation loop (on the left) and a pure screw dislocation line (on the right). The dislocation lines are shown in black along with the smectic layers defined by $\cos\phi=1$.}
\end{figure}

\begin{figure}
\centerline{\includegraphics[width=0.5\textwidth]{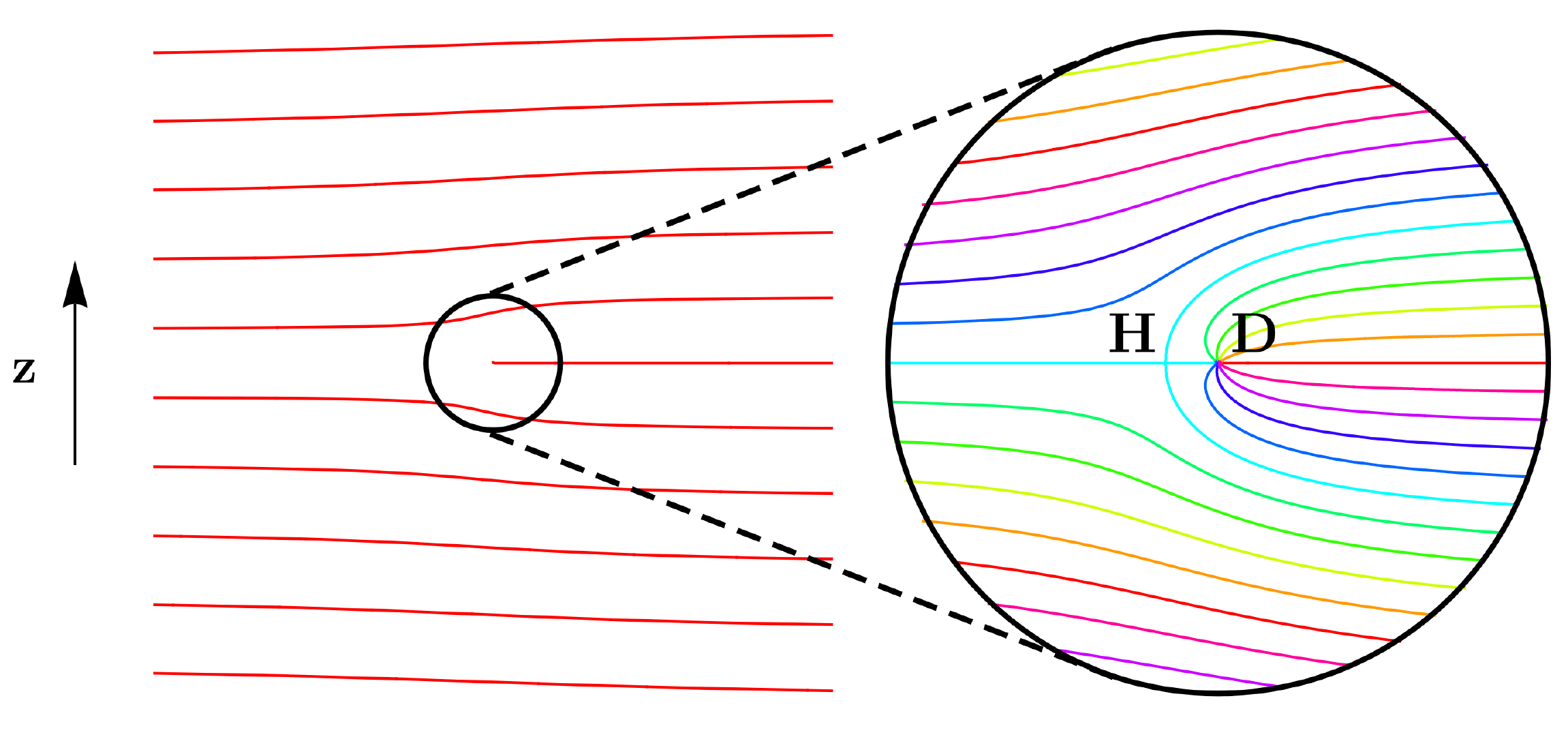}}
\caption{\label{fig:HD} We consider the layer structure of a two-dimensional edge dislocation. The layers corresponding to $\cos\phi=1$ are depicted in red. The inset shows the detailed structure of level sets of $\phi$ near the singularity at $\bf D$. Different level sets are depicted in successive colors, red$\rightarrow$yellow$\rightarrow$green$\rightarrow$indigo$\rightarrow$red.  The dislocation adds an extra complete set of level sets indexed from $0$ to $2\pi$ (mod $2\pi$).   Because the layer normals eventually point along the $\hat z$ axis, a hyperbolic point ($\bf H$) arises, connecting to the dislocation $\bf D$. Four rays, in pale blue, emanate from the hyperbolic point $\bf H$, while three of these rays escape out to infinity.  One of the rays just connects $\bf H$ and $\bf D$.  Note that since the hyperbolic point is not in the singular set of the sample, there is an unambiguous value of $\phi$ at that point, the value that we assign to $\Upsilon$.  This picture does not change qualitatively for a mixed dislocation with some screw component.  As long as it is not pure screw, the layers go off to left and right infinity to their periodic ground state.
}
\end{figure}

The general problem includes topological defects in the sample that require us to specify boundary conditions on the defect sets $\Sigma$.  It will best serve us to specify the following boundary conditions.  Around each defect we specify a winding and the layer normal, $\bf N = \nabla\phi/\vert\nabla\phi\vert = \hat z$, at infinity. Note that (\ref{eq:fe}) sets the layer spacing through energetics, but not topology.  
For concreteness, we can begin by considering an edge dislocation loop in the $xy$-plane of radius $r_0$ (depicted in Fig.~\ref{fig:pure_dislocations} as the black loop).  A phase field with the requisite topology and boundary conditions is
\begin{equation}\label{eq:phie}
\phi_{e} = z + \tan^{-1}\left(\frac{z}{r_0-r}\right).
\end{equation}
At infinity $\nabla\phi_e=\hat z$ implying $q=1$.  If we want a field with a different value of $q$, we write  $\phi_{e,q} = \phi_e + (q-1)z$.  As long as $q>0$ we have a continuous family of solutions depending on $q$ that all satisfy the specified boundary conditions.  We will exploit this homotopy later, but at this point we return to $\phi_e$.  
Note that this function is not smooth along the $\hat z$-axis, but it contains the topology we need (a smoothed out
version of it can be readily defined at the expense of functional simplicity and this has been done to obtain Fig.~\ref{fig:pure_dislocations}).  Specifically, on the dislocation the value of $\phi_e$ is undefined and, moreover, $\oint_\gamma d\boldsymbol{\ell}\cdot\nabla\phi_e = -2\pi$ for any contour $\gamma$ that circles the defect line once. It is worth pausing and investigating the detailed structure of this phase field.  Recall that both singularities and zeroes of $\nabla\phi$ correspond to disclinations in the smectic where the layer normal ${\bf N}\equiv \nabla\phi/\vert\nabla\phi\vert$ is undefined \cite{chen09};
in cylindrical co\"ordinates $(\hat r,\hat \theta,\hat z)$, 
\begin{equation}
\nabla\phi_{e} = \left[\frac {z}{ z^2 + (r-r_0)^2},0,1-\frac{r-r_0}{z^2 + (r-r_0)^2}\right].
\end{equation}
Near the singularity at $z=0$ and $r=r_0$, we expand in $\delta=r-r_0$ to find ${\bf N} \sim [z,0,-\delta]/\sqrt{z^2+\delta^2}$, a $+1$ disclination; near the zero at $z=0$ and $r=r_0+1$ we expand in $\delta=r-r_0-1$ and find ${\bf N}\sim[z,0,\delta]/\sqrt{z^2+\delta^2}$, a $-1$ disclination!  Together these create the standard dislocation as shown in Fig.~\ref{fig:HD}.  Were we to embellish (\ref{eq:phie}) to make a dislocation of strength $2\pi m$ (charge $m$ with $q=1$), we would find that the two disclinations were precisely $m$ apart, in concordance with the standard discussions of disclination dipoles \cite{halpnels}.  Similarly, were we to change the value of $q$ from $1$, the second disclination would move to $r=r_0+1/q$. It follows that deformations through the smooth family can bring the $-1$ disclination arbitrarily close to the $+1$ disclination.

The $-1$ disclination, which we will refer to as the hyperbolic or saddle point, is special to smectics.  It occurs because $\nabla\phi\rightarrow q\hat z$ at infinity: just as our special choice $\phi_e$ broke into a disclination dipole so too will any edge dislocation phase field.  Consider Fig.~\ref{fig:HD};  in the neighborhood of $\bf D$, the singularity, we have the topology of a $+1$ disclination.  Does $\nabla\phi$ have a zero?  Yes!  To see this, note that $\nabla\phi$ must point along $\hat z$ on two lines that emanate from $\bf D$ and go off to left and right infinity.   Since it diverges at $\bf D$, at a point close enough to the left of a positive dislocation $\nabla\phi$ is arbitarily large and negative while at left infinity $\nabla\phi=\hat z$.  It follows that somewhere along the line going to left infinity $\nabla\phi$ must vanish -- the $-1$ disclination.  The saddle  point is at the intersection of two asymptotic lines that all have the same value of $\phi$ or, equivalently,  four rays emanating from it.  The long distance structure of the dislocation has three rays going off to infinity, one of them arising from the extra (or missing, depending on whether you are an optimistic or a pessimist) layer.  The fourth ray must terminate on $\bf D$ creating a disclination dipole ``pitchfork'' as in Fig.~\ref{fig:HD}.

\section{Distinguishing Screw from Edge Dislocations}

To put the previous discussion into perspective, consider the canonical example of a screw dislocation (depicted by the helicoid in Fig.~\ref{fig:pure_dislocations}):
\begin{equation}\label{eq:screw}
\phi_s = z - \tan^{-1}\left(\frac{y}{x}\right).
\end{equation}
We immediately see that it does not have any component that cancels off the background $\nabla\phi=\hat z$ and therefore it does not decompose into two disclinations.
What about dislocations with a mixed edge/screw character? Let the tangent vector of the dislocation line at some point be $\boldsymbol{\tau}$.  If we take a plane perpendicular to $\boldsymbol{\tau}$ then, as long as $\boldsymbol{\tau}$ is not parallel to $\hat z$ the intersection of the smectic field with the plane shares the same topology as depicted in Fig.~\ref{fig:HD}.  The hyperbolic point is no longer a location where $\nabla\phi=0$ since $\nabla\phi$ will have components out of the plane.  However it is still a feature of the two-dimensional slice: define $\tilde\phi$ to be the restriction of $\phi$ to the transverse plane and let $\nabla_{\!{\scriptscriptstyle\perp}\!}$ be the gradient operator in that plane.  Then $\nabla_{\!{\scriptscriptstyle\perp}\!}\tilde\phi$ vanishes somewhere by the same argument as before.  Note from simple geometry that the magnitude of $\nabla_{\!{\scriptscriptstyle\perp}\!}\tilde\phi$  at infinity gets smaller as the angle of $\boldsymbol{\tau}$ with the $\hat z$ axis decreases and the saddle moves farther and farther away from the dislocation: this is why $\boldsymbol{\tau}$ cannot be parallel to $\hat z$ - the saddle moves off to infinity as in the case of the pure screw dislocation line of Eq.~(\ref{eq:screw}).
Depending on the limit, however, it can escape out to infinity in any direction.  If we put the smectic in, for instance, a cylinder ${\bf D}^2\times\mathbb{R}$ then we have an ambiguity arising from the changing boundary condition on the cylinder at infinity.  
Finally, note that in the presence of other dislocations it is possible for the argument predicting hyperbolic points to fail; zeroes of $\nabla\phi$ are not conserved, only singularities are.  The hyperbolic point associated with any particular dislocation may disappear into the other dislocation.  However, by increasing $q$ through our smooth set of smectics we can, as we argued, isolate the zero arbitrarily close to the dislocation. We will return to this point later.

So we see that a screw dislocation is different from an edge -- it has no hyperbolic point.  Can we make this more precise?  Note that we are required  to specify a fixed sample topology with boundary conditions in order to solve the Euler-Lagrange equations, for instance.  If we take a finite sample with fixed boundary conditions $\nabla\phi =q\hat z$ then we can add the point at infinity and consider the problem on ${\bf S}^3$.  However, even this is problematic from the point of view of $\phi$ -- there is a defect at infinity, with the topology of an ``electric dipole'' in $\nabla\phi$.  One might recall that Morse theory forbids such structures \cite{poenaru} in a smectic.  It is instructive (and easier to visualize) to consider the situation for a two-dimensional smectic.  Suppose we want to have a defect free smectic on $\mathbb{R}^2$;  there are an infinite number of layers and there is some equally-spaced ground state at infinity, a simple periodic structure.  If we want to compactify this to $\bf{S}^2$ then we must squeeze all the layers together at infinity or, in other words, on the sphere we want to have a set of non-crossing layers except, perhaps at infinity, the North pole.  The Euler character of the sphere being two implies that there must be a $+2$ winding defect at infinity.  In one dimension up the Euler character of the sphere is zero and this implies that $\nabla\phi$ must be wrapped up at the North pole in a defect with index zero. A direct calculation shows that this is what happens with the aforementioned ``electric dipole'' texture.

Because of this complication we will restrict ourselves to finite regions $M\subset\mathbb{R}^3$ with the rigid boundary condition $\nabla\phi = q\hat z$. Note that the Euler-Lagrange equations that follow from minimizing the free energy are {\sl fourth}-order requiring three boundary conditions.  This boundary condition supplies three as do the windings around each defect -- the phase change and the planes in which the phase changes.   As a consequence of the rigid boundary conditions on $\partial M$ it is impossible for a hyperbolic point in a plane {\sl not perpendicular} to $\hat z$ to pass through the boundary: if the plane is normal to $\boldsymbol{\tau}\!\!\not{\!\parallel} \hat z$ then at the hyperbolic point, $\nabla\phi\!\!\parallel\!\!\boldsymbol{\tau}$.  If, on the other hand, $\boldsymbol{\tau}\!\!\parallel\!\!\hat z$ then the argument above shows that the hyperbolic point is not required and there is no obstruction at $\partial M$.  This implies that a piece of edge dislocation {\sl cannot} be twisted into a screw dislocation while maintaining the boundary conditions.  {\sl The boundary conditions distinguish screw from edge dislocations through the topology of the in-plane phase field $\tilde\phi$.}  In a finite sample this is a difference between a edge-like and screw-like dislocation.  The former have their hyperbolic points inside the sample, the latter outside.  If we think of the divergence-hyperbolic point pair as the generalization of the core of the pure edge defect, then screw-like dislocations have cores that are larger than the sample!  This is the physical difference between the two types of defect -- localized versus delocalized cores. For an edge defect to become a screw defect (or {\sl vice versa}), we must relax the boundary conditions to let the hyperbolic point through. We will call a dislocation with no pure (100\%) screw component a ``some-edge'' defect since all parts of it have some edge component. Since we are considering only ambient isotopy of the dislocations, we may keep them all at a minimum distance from each other and thus we conjecture (in {\sl lieu} of a proof) that the hyperbolic points cannot swap among dislocations.

\section{Tethering of Dislocation Loops}
\subsection{Example: An Edge Loop around a Screw Dislocation}

A consequence of this is a  topological conservation law; a defect loop with no screw component is topologically tethered to the other defects that it links.  Because every point on the dislocation has a unique partner hyperbolic point, we can assign a value of $\phi$ to each point on the singularity.  This is not possible in a superfluid since there is no general way to assign a value of the phase on a vortex -- all values of the phase exist on the singularity.  It is only the natural framing of a some-edge dislocation by its hyperbolic point that allows this.  The  vortex in a superfluid is akin to a section of pure screw defect: $\nabla\phi\rightarrow 0$ in a superfluid because $\phi$ goes to a constant on the boundary while $\nabla_{\!{\scriptscriptstyle\perp}\!}\phi\rightarrow 0$ for a screw dislocation by the geometry of the smectic layers.

\begin{figure}
\begin{center}
\includegraphics[width=0.3\textwidth]{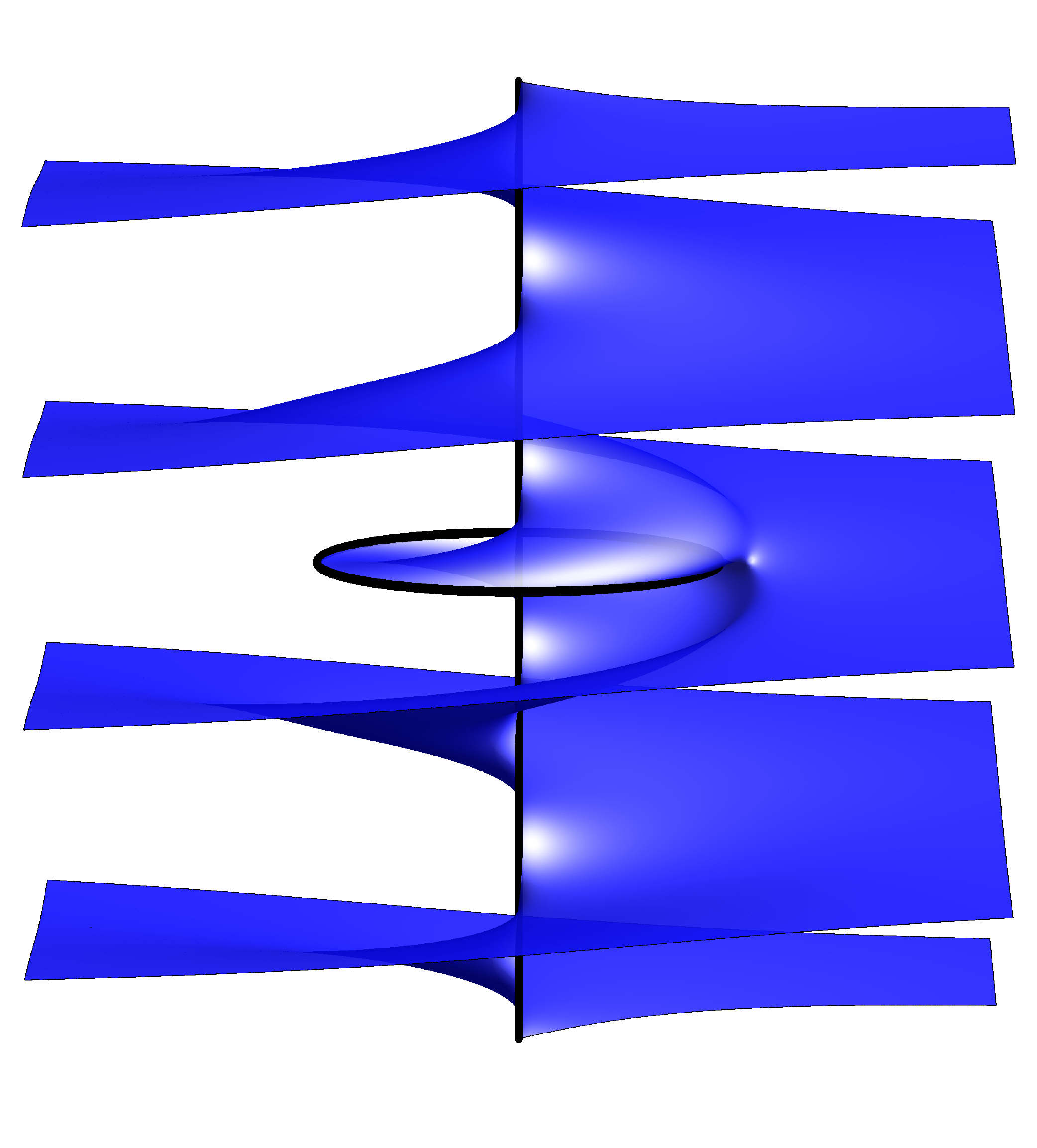}
\includegraphics[width=0.36\textwidth]{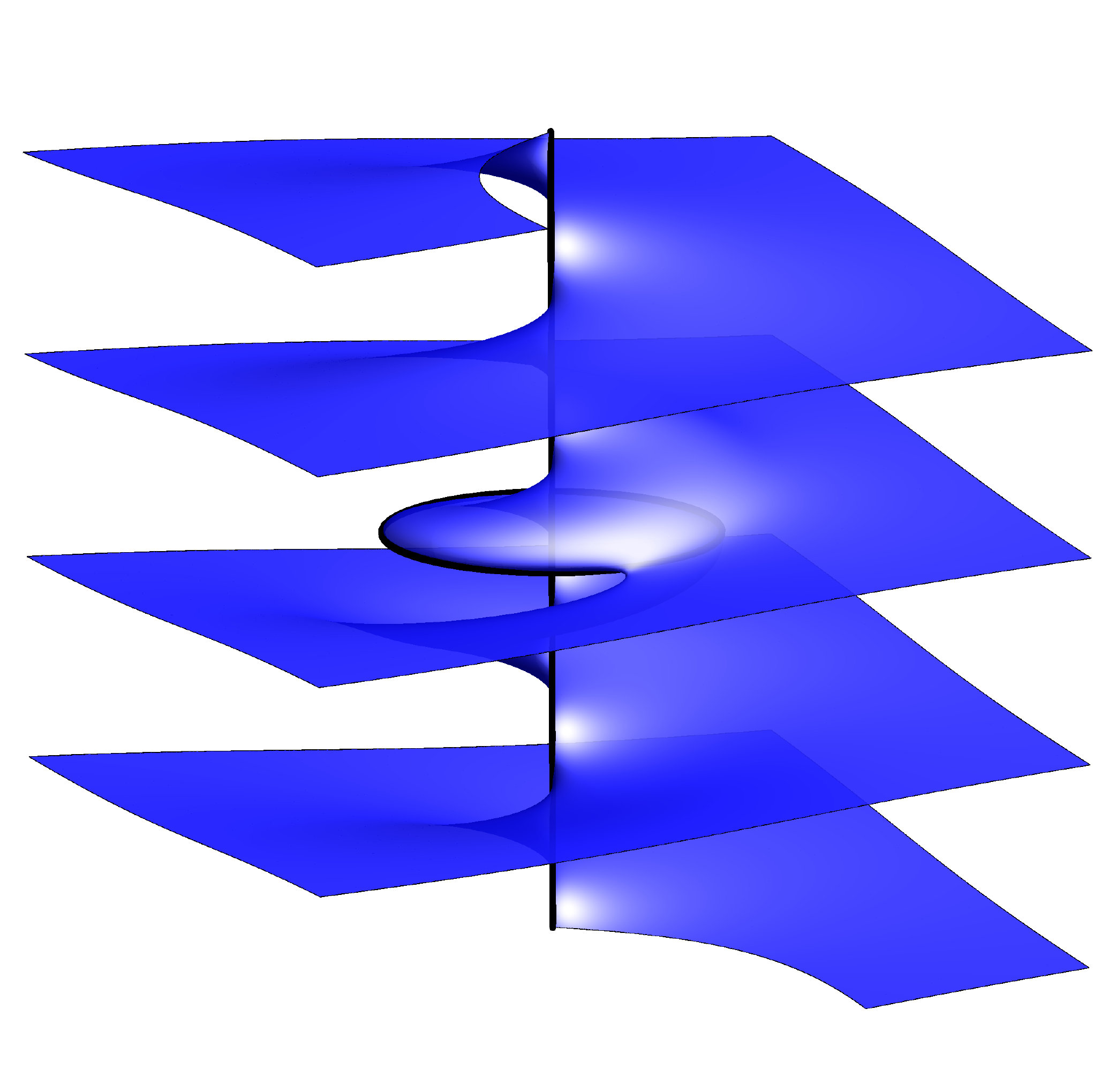}
\caption{\label{fig:edge+screw} A pure edge dislocation loop pierced by a pure screw dislocation line. The figure shows, from two different points of view, the dislocation lines (in black) along with the smectic layer defined by $\cos\phi=1$. See SI for a movie of the full $2\pi$ rotation around the screw dislocation. See Movie1 in the SI for more points of view of this geometry.}
\end{center}
\end{figure}

\begin{figure}
\centerline{\includegraphics[width=0.3\textwidth]{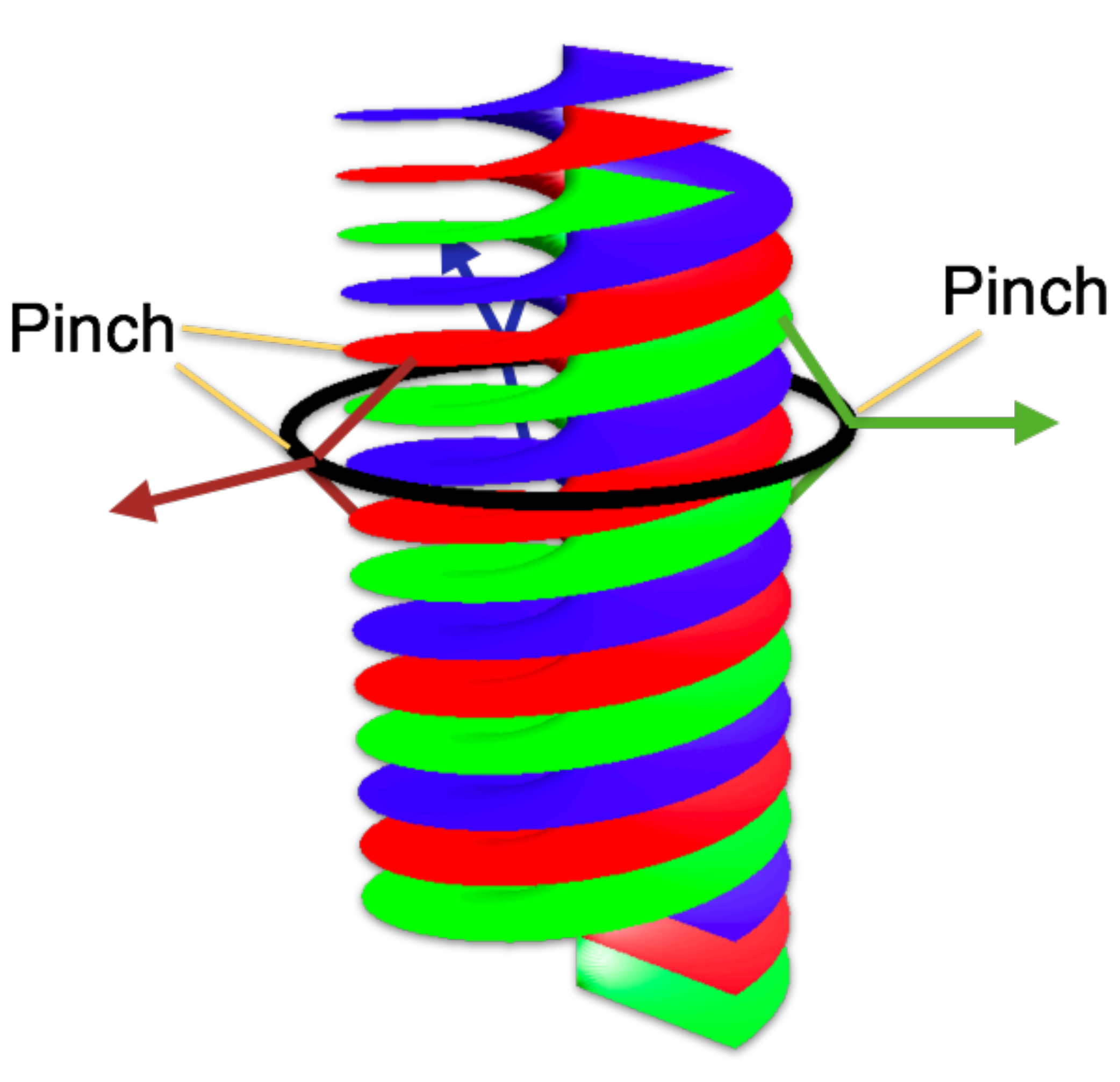}}
\caption{\label{fig:pinch} The geometry of an edge loop threaded by a screw dislocation.  Inside the loop we have a standard screw dislocation.  At the edge loop we pinch together two consecutive layers of the same color on to the loop.  As we go around the edge loop, the local color changes and goes through all $2\pi$ of the phase.  Outside the edge dislocation loop we have another helicoidal structure with one less layer.  
}
\end{figure}

\begin{figure*}
\centerline{\includegraphics[width=0.9\textwidth]{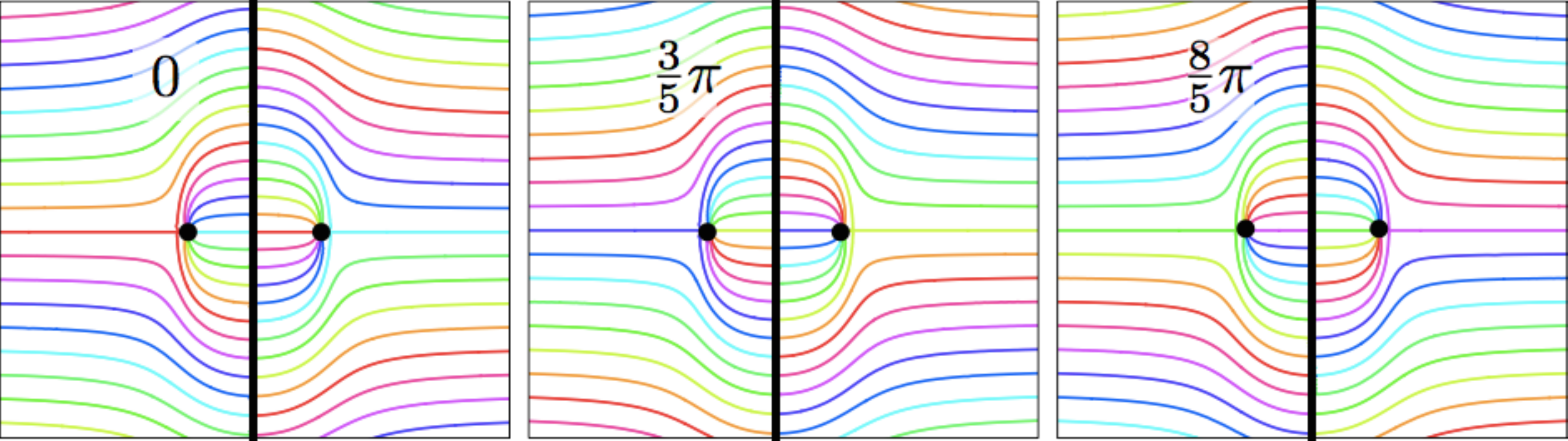}}
\caption{\label{fig:rotate} {Edge dislocation loop pierced by a screw dislocation line (as in Fig.~\ref{fig:edge+screw}). The figures show the smectic layers on vertical planes containing the screw for different values of the cylindrical angle $\theta$. Different level sets of $\phi$ (mod $2\pi$) are depicted in successive colors, as in Fig.~\ref{fig:HD} . Notice that, as $\theta$ increases, the hue of the layer connecting the edge to infinity also increases. As a result, $\Upsilon$ changes by $2\pi$ after a complete turn around the screw. See SI for a movie depicting views of this structure from varying viewpoints. See Movie2 in the SI for a complete tour through all the angles.}
}
\end{figure*}
Expanding on this, consider a defect in a superfluid.  One might think that a closed vortex line would accumulate the winding imposed by the vortices it surrounds.  However, if we try to measure the winding on the vortex we run into a problem -- the value of the phase is, by definition, undefined on the defect.  We could frame the vortex loop, but there is no well-defined way to do this.  We could, for instance, push off the vortex line along a direction of some fixed value of $\phi$.  But how far?  Two different experiments could get two different answers by pushing off different distances.  An unambiguous approach that preserves the helicity is described by Moffat and Ricca \cite{moffat2}.  There they replace each vortex line with a vortex ribbon of fixed width, introducing at the outset a preferred framing.  This geometric replacement of the lines leads to a truly conserved helicity (in ambient isotopy of the {\sl ribbons}) when C\v{a}lug\v{a}reanu's theorem relating twist, writhe, and link \cite{cala,hannay,rmp} is included in the analysis.
This is the essential difference -- some-edge dislocations have a different framing allowing us to define a different invariant that comes from the phase field itself.  

We define $\Upsilon$ on the dislocation line $\gamma$ as the value of $\phi$ on the corresponding hyperbolic point.  The change in phase integrated around $\gamma$ must be a multiple of $2\pi$:
\begin{equation}
\Delta\Upsilon = \oint_\gamma d\Upsilon \in 2\pi\mathbb{Z}.
\end{equation}
An interpretation of $\Upsilon$ is possible in the smectic case that fails in the superfluid situation: $\Upsilon$ is the value of the level set at the boundary that comes into the sample and joins the edge dislocation at that point.   It is only because the value of $\phi$ grows linearly  (in the $\hat z$ direction) on the boundary that we can make this identification.  Thus, despite the fact that $\phi$ is not defined on the dislocation there is still a special value of $\phi$ associated with a boundary condition.  For each value of $\Upsilon\,\textrm{mod}\,\, 2\pi$
we have a different, but equivalent, set of level sets $\phi-\Upsilon\in 2\pi\mathbb{Z}$.  On an edge dislocation there is precisely one value of $\Upsilon$ that attaches to it.  

In order to make these ideas concrete we consider the example of an edge dislocation loop pierced by a screw dislocation line:
\begin{equation}\label{eq:edge+screw}
\phi = z - \tan^{-1}\left(\frac{y}{x}\right) + \tan^{-1}\left(\frac{z}{r_0-r}\right).
\end{equation}
The geometry of this configuration is depicted in Fig.~\ref{fig:edge+screw}. Fig.~\ref{fig:pinch} shows one way to understand how it comes about. We can also unwrap this three dimensional geometry to create Fig.~\ref{fig:rotate} where we show a cross section of an edge loop threaded by a single screw dislocation for various values of the cylindrical angle $\theta$. In the leftmost frame where $\theta=0$ we see that the red phase value comes in from the left boundary while cyan comes in from the right.  The two colors are apparently $\pi$ apart in hue.  However, when we rotate to $\theta=3\pi/5$ the left and right colors change -- these are coming from different values of $\Upsilon$ on the boundary.  Similarly in the rightmost frame where we rotate by $8\pi/5$.  Each point on the edge loop has a different value of $\phi$  and, as we go around the screw defect the value of $\Upsilon$ winds by $2\pi$.  This geometric identification with the boundary condition is impossible for vortices in the superfluid since $\phi$ becomes constant on the boundary and, for the same reason, fails for a pure screw dislocation.  If an edge dislocation in the smectic were allowed to distort into a pure screw then the phase accumulated on the edge could slip out because $\Upsilon$ can not be defined there.  Fortunately, this cannot happen in our finite sample, as we noted.  Were we to try to take the no boundary limit by using ${\bf S}^3$, then the phase slips as the hyperbolic point moves out to infinity and winds around the ``electric dipole'' defect at infinity.  Thus the separation of pure screw components from defect segments with some edge component is not only necessary from the point of view of the boundary conditions but it is essential for the conservation of this new linking invariant.

\begin{figure}
\centerline{\includegraphics[width=0.5\textwidth]{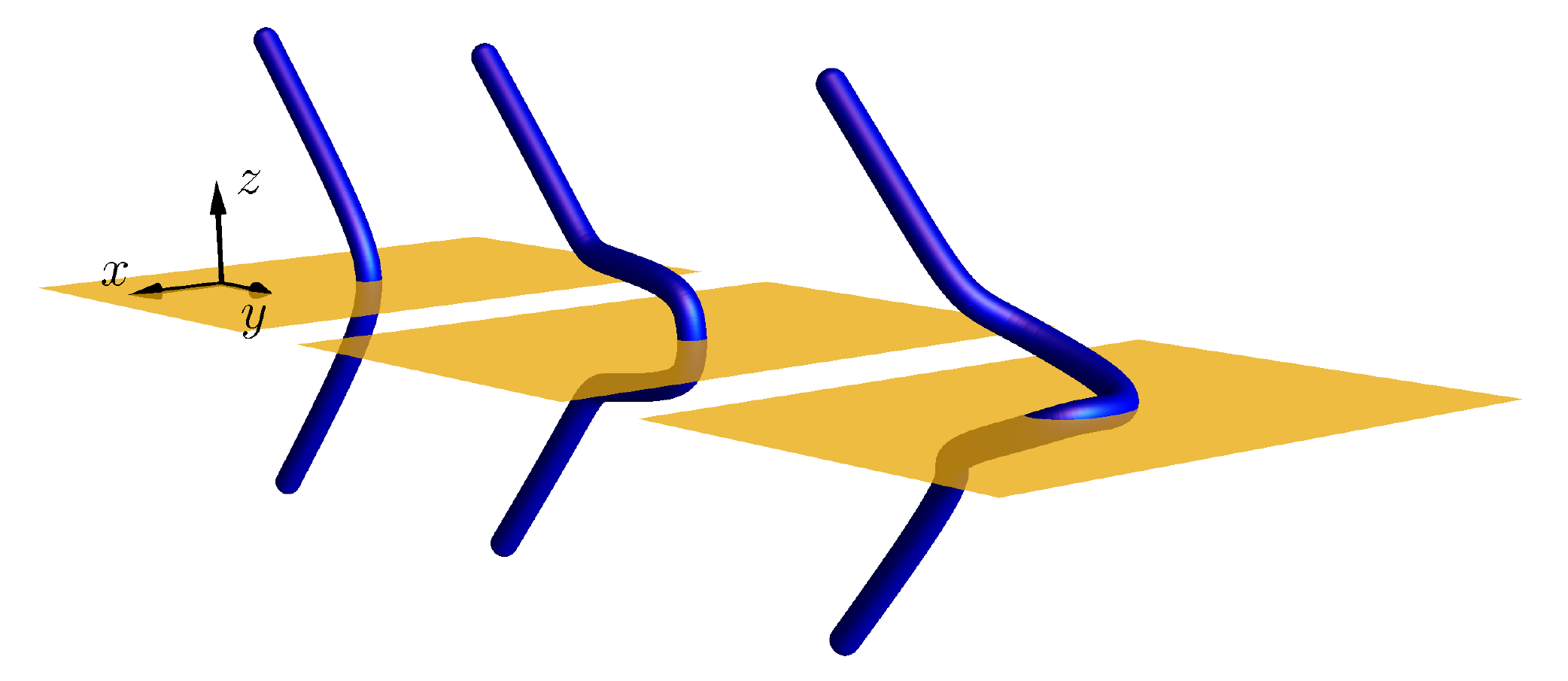}}
\caption{\label{fig:reg} A regularization of a point wherein the dislocation line is vertical. The original line (left panel) is pinched near that point (central panel) and then twisted so that no portion of it points along the $z$-direction (right panel).}
\end{figure}

We could extend all this to arbitrary dislocation loops so long as we regulate the portions of the dislocation line that are purely screw.   In order to do so, we pick a maximum slope $m$ and, whenever the curve exceeds that, we replace it with another curve with slope not exceeding $m$. This can be realized, for instance, with a helix or with the twisted curve of Fig.~\ref{fig:reg}.  To remove self-winding we insist that the approximation to a vertical section {\sl not} wind around the actual dislocation.  This is just choosing the measuring loop to not link with the actual dislocation and can be corrected, if we wanted to, by subtracting off the linking of the two loops.  We can then find all the hyperbolic points by choosing large enough $q$ and calculate $\Upsilon$.  Since this prescribes a value of $\Upsilon$ for the original dislocation, we can calculate the winding number.  Moreover, since the value of $\Upsilon$ arises from measuring $\phi$ in $M\setminus\Sigma$, the winding of $\Upsilon$ is {\sl precisely} the winding imposed by any defect that links the loop of interest.  Thus a dislocation loop $\boldsymbol{\gamma}$ that is unlinked and unknotted, the ``unlink'', has $\Delta\Upsilon=\oint_{\boldsymbol{\gamma}} d\boldsymbol{\ell}\cdot\nabla\phi=0$ while a loop linked by loops with charges $n_i$ will carry a charge of $2\pi \sum_i n_i$.  
In fact, unlike the original helicity invariant which has trouble because of the diagonal self-linking terms \cite{moffat2}, this invariant is precisely the sum of the linking numbers -- there is no self-interaction term since $\Upsilon$ cannot pick up a winding from its own defect (which is why we approximate screw sections by curved segments that do not wind around the actual defect!). 
Note that here the framing is provided automatically by the hyperbolic points and thus differs from the helicity. It also differs from helicity in that the mutual linking among the links that pierce the measuring loop do not contribute to the invariant. Each loop independently has its own well-defined topological linking with the other loops.  

In closing this section, it is amusing to study an edge dislocation loop pierced by a screw dislocation from the perspective of the Volterra theory of defects.  Consider an isolated screw dislocation at the origin and a dislocation loop that {\sl adds} an extra disc of smectic away from the origin.  We can slide this extra disc around keeping it between the two layers that bind it.  Though technically this motion is a dislocation climb since the Burgers's vector is perpendicular to the motion, it is especially low energy in the smectic because the motion is perpendicular to any periodicity.  As the dislocation loop slides up to the screw dislocation, half of it must go ``up the spiral staircase'' and the other half must go down the staircase.   If we insist that the layers do not cross there is some sort of tear in the original extra disc.  This introduces an extra structure in the smectic texture.  Note that we must compare bananas to bananas: we do not consider an edge dislocation spiralling up and around a screw defect ``linked'' with the screw.  The edge dislocation must close back on itself.  It is through our phase field approach that we can make this construction precise to show that there is, in fact, a topologically stable difference between the link and unlink.  Note that in all our phase fields and in Figs. 5-7 the singularity of the edge dislocation remains on a closed loop confined to the plane $z=0$.  Though there is a complex, helicoidal-like structure shown in Fig.~\ref{fig:edge+screw}, the edge dislocation, in black, is a closed circle.  {\sl Beware the Volterra construction!}

\begin{figure}
\centerline{\includegraphics[width=0.5\textwidth]{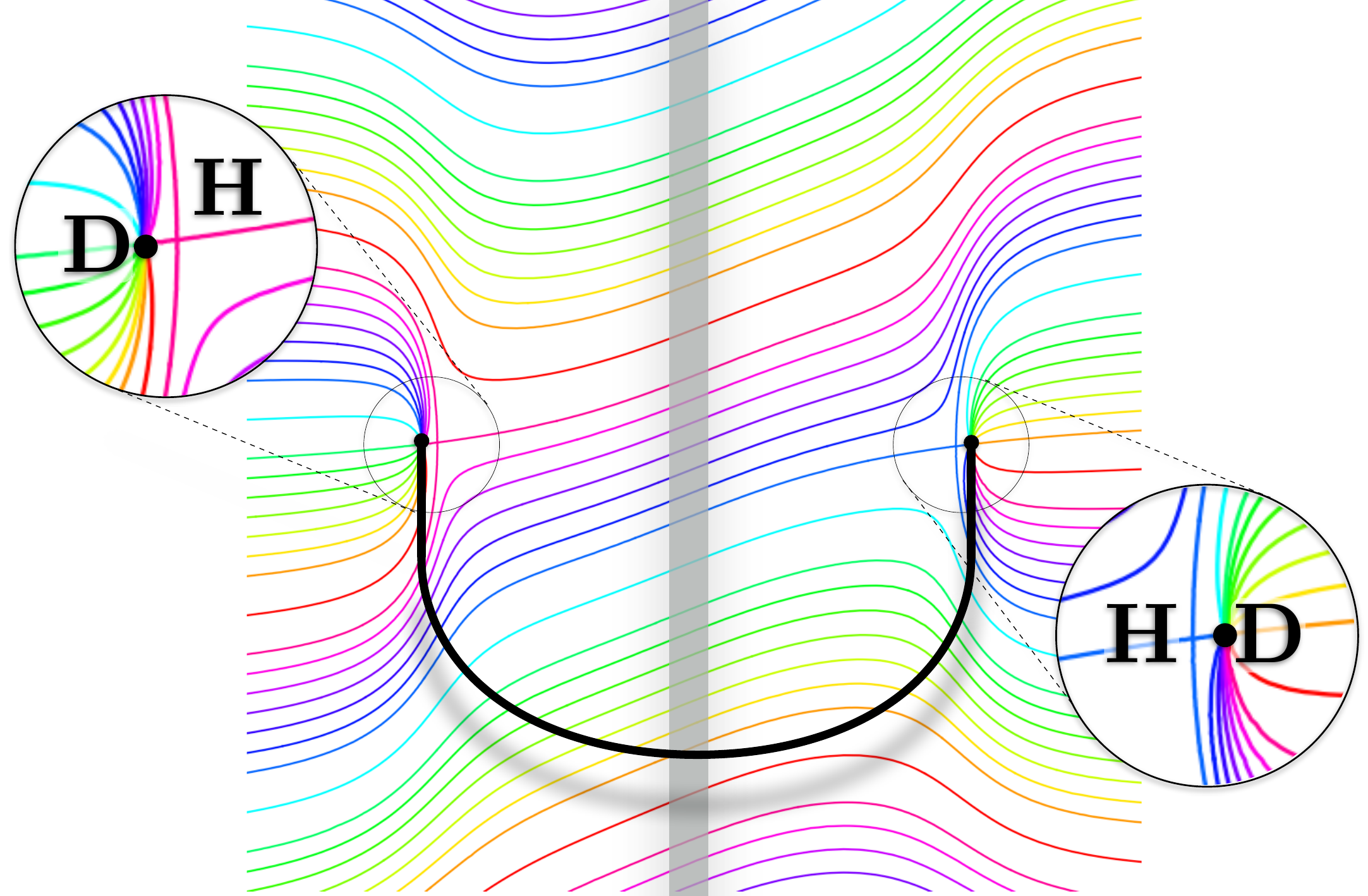}}
\caption{\label{fig:tether}  The black line represents the core of a single line defect that has wrapped around a pure screw dislocation that sits vertically in the plane of the page along the $\hat z$-axis, in grey.  We show the level sets of constant phase in an $xz$-plane behind the plane of the page.  The black edge dislocation begins and ends even further behind the plane of the page and comes out and loops around the screw dislocation in front of the page.  We now see that the lines of constant phase have an unremovable jog in them that cannot be eliminated without introducing more defects that collapse the lines together.  This is the tell-tale topological tether that cannot be removed in the smectic.  In a superfluid the different colored lines would wrap around and find each other since $\phi$ goes to a constant at the boundaries.  If the edge dislocation did not wrap the screw defect, the black line would come straight out of the page.  There would be, in that case, no phase difference to measure since there would be no corresponding singularity on the opposing side of the screw.  We can only measure phase differences and, to do so, we need a singularity on either side of the screw dislocation.
}
\end{figure}

\subsection{General Loops and $q$-Homotopy}

When there are multiple defects the unique pairing with hyperbolic points can disappear -- though the singularities in $\nabla\phi$ cannot disappear, zeroes  in $\nabla\phi$ may not behave well when two dislocations get close.  It is only the condition on the boundary $\nabla_\perp\phi\ne 0$ that guarantees the hyperbolic point and another defect can spoil this.  For instance,  consider a parallel pair of edge dislocations at the same $z$ height arranged with their hyperbolic points pointing towards each other, a dislocation/anti-dislocation pair with the extra layers on the ``outside''. There is no argument that ensures that there are zeroes in $\nabla\phi$ and that those would remain close to each singularity.  Indeed, as they get closer the hyperbolic points will first coalesce and then become delocalized.  Letting the distance between them be $d$, the phase field would be 
\begin{equation}
\phi_d = z-\tan^{-1}\left(\frac{y}{z}\right) + \tan^{-1}\left(\frac{y+d}{z}\right)
\end{equation}
and the zeroes of $\nabla_\perp\phi$ are located at $y=-\frac{d}{2} \pm \frac{1}{2}\sqrt{d(d-4)}$, $z=0$ for $d\ge4$ and $y=-\frac{d}{2}$, $z=\pm \frac{1}{2}\sqrt{d(4-d)}$ for $0<d\le4$. It follows that when $d\in(0,4]$ the hyperbolic points are equidistant from the singularities and cannot be attributed to a unique dislocation -- they really belong to the dislocation pair. (Note that when $d<0$ the  defect pair is in the opposite orientation with the hyperbolic points on the outside.)  Without a unique prescription of a hyperbolic point we are at a loss to define $\Upsilon$.  This is where the equivalence between states with different values of $q$ comes in.

If we know $\Upsilon_q$ for some choice of $q$ then $\Upsilon_q = (q-1)z +\phi=(q-1)z+\Upsilon$.  This allows us to define $\Upsilon$ even when dislocations interfere, allowing us to assign a phase to the dislocation.  For instance, we see that $\phi_{d,q}$ gives us zeroes at
$z=0$ and  $y= -\frac{d}{2}(1\pm \sqrt{1-4/(qd)})$.  Thus for any fixed $d$ we can always make $q$ large enough to resolve the zeroes.    Finally, for any $\phi_q$ the boundary conditions are unchanged -- the winding is specified around the defects and $\nabla\phi\rightarrow\hbox{constant}$ on the boundary.  
So, if we start with an initial configuration of isolated defects then there is a minimum distance between them.  By choosing large enough $q$ we can resolve all their hyperbolic points and assign the value of $\Upsilon$ to the some-edge portions of a dislocation.

We finally return to the original question, ``do dislocation loops topologically link?''  More precisely, ``under ambient isotopy does the tether between two crossing defects carry non-trivial topology?''  Yes!  As shown in Fig.~\ref{fig:tether} the topology of the phase field is twisted between two parts of a dislocation line that has encircled a screw defect (sitting vertically above the plane of the page).  Because the colored lines all go off to the boundary on the right and left, the kink cannot be undone without connecting and disconnecting some of the lines or, in other words, introducing more topological defects.  But this could locally be the phase field of a superfluid, so isn't the superfluid tethered? No! Since the superfluid phase goes to a constant on the boundary, all of the constant-$\phi$ lines must loop around and end on themselves: this follows from the construction of the level sets.  Without any loss of generality, we can take the value of $\phi$ to be $0$ on the boundary.  No other level set can cross the $\phi=0$ set and it follows that any level set inside the sample stays in the sample and the only possible boundary of a level set is a dislocation.  Now consider, for instance, the violet line in Fig.~\ref{fig:tether} that starts below the two dislocations on the left and ends above the two dislocations on the right.  In the superfluid we can bring the violet end on the left  into contact with the violet end on the right because we can always homotope the actual boundary condition into a {\sl violet} loop going around the whole sample, that is, either the violet line must reconnect with itself or end on the boundary which, itself, is a closed loop.  This closed loop can then be shrunk to surround only one of the defect lines -- a superfluid vortex. We can use this to unwind the violet phase line from the two dislocations.  The smectic, with its fixed winding on the boundary, cannot participate in this homotopy without making $\phi$ constant somewhere on the boundary -- new defects that we did not ``order.''  Equivalently, we note that a location where $\nabla\phi=0$ is {\sl not} a defect in a superfluid, but is a disclination in the smectic.  The standard arguments show that the disclination loop of negative charge associated with the dislocation can not end.  Because the phase is well defined on that disclination, it is impossible to remove the winding in $\Upsilon$.  The ``negative disclination'' can disappear in the superfluid -- it wasn't there at the outset.

\subsection{Knots: Another Interpretation of $\Delta\Upsilon$}

Part of the restrictions of the previous sections can be removed just by introducing a little more machinery along the lines of the elegant work of Dennis and Hannay \cite{hannay}.  The smectic offers us two different ways of framing a dislocation loop with everywhere nonvanishing edge component.
Consider a dislocation line $C$ and define the curve $C_0$ as the push off of $C$ along the layer $\phi=0$ (we can, of course, make $C_0$ arbitrarily close to $C$). Let $C_h$ be the hyperbolic line associated with $C$ (we can also make it arbitrarily close to $C_0$ by making $q\to\infty)$. Now, by C\v{a}lug\v{a}reanu's formula \cite{cala,rmp}, we have $\Lk(C_0,C)=\Tw(C_0,C)+\Wr(C)$ and $\Lk(C_h,C)=\Tw(C_h,C)+\Wr(C)$ where $\Lk(C_1,C_2)$ is the Gauss linking number of the two curves $C_1$ and $C_2$, $\Tw(C_1,C_2)$ is the twist of the vector pointing from $C_1$ to $C_2$ around $C_1$, and $\Wr(C)$ is the writhe of the curve $C$, a geometric quantity. 
We can eliminate writhe to find
$\Lk(C_0,C)-\Lk(C_h,C)=\Tw(C_0,C)-\Tw(C_h,C)$.   But consider the value of $\phi$ on the hyperbolic point.  It winds {\sl precisely} because the zero of the phase field twists about $C$ in response to the other dislocations it corrals.  In the geometry we have so far considered, the hyperbolic line does not twist around $C$.  Thus $\Tw(C_0,C)-\Tw(C_h,C)=\Delta\Upsilon/2\pi$ (recall that twist has a denominator of $2\pi$ by convention).  
Therefore
\begin{equation}
\Lk(C_0,C)=\Delta\Upsilon/2\pi+\Lk(C_h,C).
\label{Ca}
\end{equation}
But now we see the problem: $\Lk(C_0,C)$ is robust while the two terms on the right hand side of (\ref{Ca}) are not.  In particular, the linking of $C$ with $C_h$ requires $C_h$ to stay in a finite region.  As we have argued, when a dislocation becomes too vertical, $C_h$ leaves the sample.  When it comes back inside it can come back on the other ``side'' of $C$ and change $\Lk(C,C_h)$.  This will then require that $\Delta\Upsilon$ compensate to keep the linking of $C$ with $C_0$ constant!  Here again, we see the difficulty with pure screw components in the dislocation -- it spoils the invariant along with the boundary conditions.\footnote{For loops that are never vertical, that is, with some edge component, we can project everything onto the $xy$ plane with the projection of $C_h$ being always at the same side of the projection of $C$. Thus, in the terminology of Dennis and Hannay \cite{hannay}, $\Lk(C,C_h)$ has no contribution coming from local crossings -- it is mainly writhe.}

This connection between the linking and $\Delta\Upsilon$ gives us, however, another interpretation:  consider a fibred knot $K$, a knot that is the boundary of a continuum of non-intersecting, space-filling surfaces each labelled by an angle on $[0,2\pi)$.  This implies an angle field $\alpha({\bf x})$ defined everywhere but on $K$.  The smectic phase field $\phi({\bf x})=z - \alpha({\bf x})$ can be constructed and can be fit to the required boundary conditions when $K$ is inside the sample.  In the absence of any other defects we can choose, for instance, the Seifert surface corresponding to the angle $\alpha=0$, $M_0$.  If we intersect this surface with an $\epsilon$-tube around $K$ then we get the curve $K_0$.  Since $M_0$ is oriented we can push $K_0$ off of $M_0$ along its normal vector and so $K_0$ {\sl does not} intersect $M_0$.  It follows that $\Lk(K_0,K)=0$.  Thus if we have a knotted dislocation with no other defects present that also has a hyperbolic line everywhere ({\sl i.e.} it always has enough edge component), $K_h$, then
\begin{equation}
\Delta\Upsilon=-2\pi\, \Lk(K_h,K).
\end{equation}
Note that although it is only necessary to have a Seifert surface to generate $K_0$, the requirement of no other defects can only be satisfied by a fibred knot\footnote{We thank Mark Dennis for suggesting the connection between dislocations and fibred knots.} -- we need a phase field $\alpha$ everywhere in space with nonvanishing $\nabla\alpha$.

\section{Conclusion}
In conclusion, we have demonstrated that, once again \cite{poenaru,chen09,trebin82}, the standard homotopy theory treatment of topological defects fails for the smectic even in the case of pure dislocations.  In addition, we have demonstrated a topological difference between the screw and edge dislocations that manifests itself geometrically -- a pure edge always splits into a disclination dipole, while a defect with pure screw character does not. A mixed dislocation still has a hyperbolic point in the transverse plane and so is more like an edge than a screw.  It is worth observing that the geometry around a dislocation with a hyperbolic point cannot be a free disclination as it will not fit into Po\'{e}naru's classification of two-dimensional critical points \cite{poenaru} -- were we to isolate the dislocation we would have a location where an infinite number of layers come together, a situation that is incompatible with smectic order as it requires infinite compression, {\sl i.e.} it is not a measured foliation.  In the case of a pure screw, however, those arguments do not apply and the dislocation can stand by itself, forcing the hyperbolic point off to infinity.
All of these observations arise from the boundary condition that the layer normal points along a constant direction and the layers have equal spacing.  This difference is what distinguishes the smectic from the superfluid where the phase field $\phi\rightarrow{\rm constant}$ on the boundary.  Whether any progress can be made by studying the fundamental group of the jet bundle that ties together $\phi$ and $\nabla\phi$ \cite{arnold} is a question for further research.  We also note that the description of crossed defects in crystals in terms of jogs and kinks \cite{hirth} might be described in a similar fashion, taking on a topological character.

\acknowledgments

We acknowledge stimulating discussions with T. Castle, B.G. Chen, M.O. Lavrentovich, C.M. Modes, J.P. Sethna, and D.M. Sussman.  We also thank T. Machon for his lucent explanation of knot theory.  This work was supported through NSF Grant DMR12-62047.   This work was partially supported by a Simons Investigator grant from the Simons Foundation to R.D.K.  R.A.M. acknowledges support from FAPESP grant 2013/09357-9.

\end{document}